\documentclass[conference]{IEEEtran}
\IEEEoverridecommandlockouts

\IEEEsettopmargin[t]{t}{0.75in}
\IEEEsettopmargin[t]{b}{1.02in}
\usepackage{textcomp}
\usepackage{caption}
\usepackage{subcaption}
\usepackage[utf8]{inputenc}
\usepackage[n,landau,notions,ff,mm]{cryptocode}
\usepackage{xcolor}
\usepackage{graphicx}
\usepackage{url}
\usetikzlibrary{shapes}
\usepackage{import}
\usepackage{xifthen}
\usepackage{pdfpages}
\usepackage{transparent}
\usepackage{balance}
\usepackage{xspace}
\usepackage{amsmath}

\usepackage{hyperref}
\usepackage{cleveref}
\usepackage{amsfonts}
\usepackage{multirow}
\usepackage{graphicx}
\usepackage{booktabs}
\usepackage{caption}
\usepackage{enumitem}
\setlist[itemize]{left=0pt}
\setlist[enumerate]{left=0pt}
\usepackage{makecell}
\usepackage{colortbl}

\usepackage{tikz}
\newcommand{\circnum}[1]{%
  \ifnum#1>0
    \ifnum#1<100
      \tikz[baseline=(char.base)]\node[
        circle,
        fill=black,
        text=white,
        inner sep=0pt,
        minimum size=0.8em,
        text width=0.8em,
        align=center,
        font=\sffamily\bfseries\scriptsize
      ] (char) {#1};%
    \else
      \textbf{?}%
    \fi
  \else
    \textbf{?}%
  \fi
}

\usepackage{threeparttable}
\usepackage{tikz-qtree}
\usepackage{pgf-umlsd}
\usepackage{dashbox}
\usepgflibrary{arrows} 
\usepackage{amssymb}
\usepackage{pifont}

\def\BibTeX{{\rm B\kern-.05em{\sc i\kern-.025em b}\kern-.08em
    T\kern-.1667em\lower.7ex\hbox{E}\kern-.125emX}}

\usepackage{acro}
\acsetup{single}

\DeclareAcronym{DeFi}{
  short = DeFi,
  long  = Decentralized Finance,
}

\DeclareAcronym{PoW}{
  short = PoW,
  long  = Proof-of-Work,
}

\DeclareAcronym{PoS}{
  short = PoS,
  long  = Proof-of-Stake,
}

\DeclareAcronym{PBS}{
  short = PBS,
  long  = Proposer-Builder Separation,
}
\newcommand{\PBS}{\ac{PBS}\xspace}

\DeclareAcronym{ePBS}{
  short = ePBS,
  long  = Enshrined PBS,
}
\newcommand{\ePBS}{\ac{ePBS}\xspace}

\DeclareAcronym{POF}{
  short = POF,
  long  = Private Order Flow,
}
\newcommand{\POF}{\ac{POF}\xspace}

\DeclareAcronym{IDM}{
  short = IDM,
  long  = Input Data Message,
}
\newcommand{\IDM}{\ac{IDM}\xspace}
\newcommand{\IDMs}{\acp{IDM}\xspace}

\DeclareAcronym{EOA}{
  short = EOA,
  long  =  Externally-Owned Account,
}
\newcommand{\EOA}{\ac{EOA}\xspace}
\newcommand{\EOAs}{\acp{EOA}\xspace}

\DeclareAcronym{MEV}{
  short = MEV,
  long  = Maximal Extractable Value,
}
\newcommand{\MEV}{\ac{MEV}\xspace}

\newcommand{\ETH}{\ensuremath{\xspace\textsf{ETH}}\xspace}
\newcommand{\tx}{$\mathsf{tx}$\xspace}
\newcommand{\txs}{$\mathsf{txs}$\xspace}

\newcommand{\inputdata}{\textsf{input}\_\textsf{data}\xspace}
\newcommand{\MEVBoost}{\textsf{MEV-Boost}\xspace}
\newcommand{\UserMod}{\textsc{UserMod}\xspace}
\newcommand{\BuildMod}{\textsc{BuilderMod}\xspace}
\newcommand{\etal}{\textit{et al.}\xspace}
\newcommand{\us}{$\mu$s\xspace}

\newcommand{\numMeaningfulTXs}{867{,}140\xspace}
\newcommand{\StartTimeData}{Jul~$30$,~$2015$\xspace}
\newcommand{\EndTimeData}{Feb~$26$,~$2024$\xspace}
\newcommand{\StartBlockData}{0\xspace}
\newcommand{\EndBlockData}{19{,}314{,}987\xspace}

\begin{document}
\title{Toxic Ink on Immutable Paper: Content Moderation for Ethereum Input Data Messages (IDMs)}

\author{
\IEEEauthorblockN{
Xihan Xiong\IEEEauthorrefmark{2}, 
Zhipeng Wang\IEEEauthorrefmark{3},
Qin Wang\IEEEauthorrefmark{4},
William Knottenbelt\IEEEauthorrefmark{2}
}
\IEEEauthorblockA{\itshape
\IEEEauthorrefmark{2}Imperial College London, United Kingdom\\
\IEEEauthorrefmark{3}University of Manchester, United Kingdom\\
\IEEEauthorrefmark{4}CSIRO Data61, Australia\\
\{xihan.xiong20, w.knottenbelt\}@imperial.ac.uk,
zhipeng.wang@manchester.ac.uk, qinwangtech@gmail.com
}
}

\maketitle

\thispagestyle{plain}
\pagestyle{plain}
\begin{abstract}

Decentralized communication is becoming an important use case within Web3. On Ethereum, users can repurpose the transaction input data field to embed natural-language messages, commonly known as \textit{Input Data Messages} (IDMs). However, as IDMs gain wider adoption, there has been a growing volume of toxic content on-chain. This trend is concerning, as Ethereum provides no protocol-level support for content moderation.

We propose two moderation frameworks for Ethereum IDMs: \textit{(i)} \BuildMod, where builders perform semantic checks during block construction; and \textit{(ii)} \UserMod, where users proactively obtain moderation proofs from external classifiers and embed them in transactions. Our evaluation reveals that \BuildMod incurs high block-time overhead, which limits its practicality. In contrast, \UserMod enables lower-latency validation and scales more effectively, making it a more practical approach in moderation-aware Ethereum environments.

Our study lays the groundwork for protocol-level content governance in decentralized systems, and we hope it contributes to the development of a decentralized communication environment that is safe, trustworthy, and socially responsible.

\end{abstract}

\vspace{1mm}
\begin{IEEEkeywords}
Content Moderation, Decentralized Communication, Ethereum, Input Data Messages (IDMs)
\end{IEEEkeywords}

\section{Introduction}
Web2 social media platforms have revolutionized communication by enabling users to share messages globally~\cite{singhal2023sok}. Similarly, decentralized communication is emerging in the Web3 space~\cite{wang2022exploring}. A well-known example is \textsf{memo.cash}~\cite{zuo2023first}, which allows users to post messages on the Bitcoin Cash (BCH) blockchain. Besides, there is another lesser-known but growing form of decentralized communication: Ethereum-based \IDMs~\cite{xiong2025talking}.  The \inputdata field in Ethereum transactions was originally designed for smart contract function calls. However, users can repurpose this field to embed human-readable content directly into transactions. This enables a new form of communication that facilitates the sharing of immutable messages across the Ethereum network.

Decentralized communication via \IDMs is growing quickly. A recent study~\cite{xiong2025talking} has detected \numMeaningfulTXs transactions containing informative \IDMs between \StartTimeData and \EndTimeData, involving $189{,}111$ wallet addresses. 
Most \IDMs were in English ($95.4\%$) and Chinese ($4.4\%$). The content spans a wide range of topics, such as social and emotional expressions, security warnings, promotions, and on-chain requests. This suggests that Ethereum is not only a transactional platform but also a growing medium for decentralized communication.

While \IDMs can be used for positive purposes, such as emotional expression and security warnings, they can also be exploited to spread toxic content. According to~\cite{xiong2025talking}, $4.3\%$ of the detected \IDMs contained toxic content. This includes profanity, threats, hate speech, sexual content, and hidden channel links. 
For example\footnote{IDM links have been omitted to avoid directing readers to harmful content.}, in transaction {0xf360...27ab}, the message contains an explicit threat, featuring violent language and personal intimidation. At one point, the sender writes:

\begin{center}
\colorbox{red!13}{
    \begin{minipage}{0.9\linewidth}
        \textit{``Your IP is being traced right now so you better prepare for the storm...I can be anywhere, anytime, and I can [redacted]\footnotemark\ you in over seven hundred ways...''}
        \raisebox{0ex}{\includegraphics[height=2.5ex]{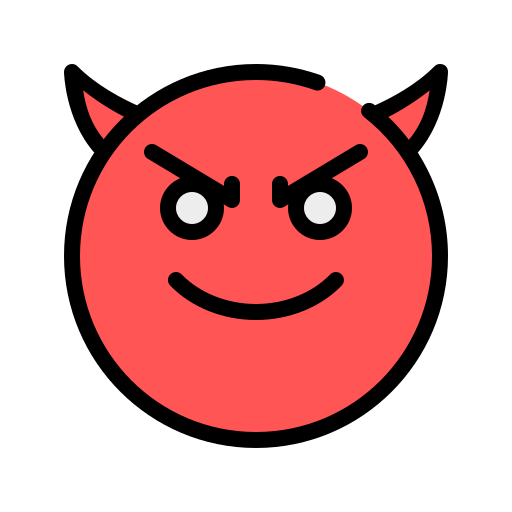}}
    \end{minipage}
}
\end{center}
\footnotetext{The original term has been omitted to avoid reproducing violent language.}

Similarly, in transaction {0x9a94...de4}, the sender issues a threat involving the receiver's entire family. Another example is transaction {0xbb68...07c}, where the message contains an explicit sexual request. More instances of inappropriate use of \IDMs can be found in our private dataset upon request.

While toxic content on Web2 platforms is typically moderated and removable~\cite{singhal2023sok}, Web3 lacks comparable mechanisms. From a governance perspective, moderating toxic content in Web3 is challenging due to the lack of centralized authority and the immutability of on-chain data~\cite{zuo2024understanding}. From a regulatory perspective, while frameworks in the EU~\cite{dsa2022} and UK~\cite{osa2023} assign clear responsibilities to Web2 platforms for managing harmful content, Web3 communication is largely unregulated~\cite{singhal2023sok}.

A key gap arises: although decentralized communication platforms serve functions similar to traditional social media, they are not subject to the same level of oversight. No mechanism for governing toxic \IDMs on Ethereum exists. However, the need to mitigate their impact is becoming increasingly urgent. On the one hand, such content can erode trust in decentralized platforms, foster hostile environments, and pose serious security risks. On the other hand, as decentralized communication expands, regulatory pressure~\cite{friedl2024decentralised} is likely to increase, making moderation both necessary and unavoidable.

Given the impact of toxic \IDMs and the pressing need for content governance, this paper explores potential approaches to moderating \IDM content within the Ethereum ecosystem. We summarize our main contributions as follows:

\begin{itemize}[leftmargin=*, topsep=0pt, itemsep=1pt]

    \item \emph{Problem Characterization}. 
    We first analyze toxic \IDMs on Ethereum and discuss their real-world impact. We then compare the legal frameworks for content moderation in Web2 and Web3. We further identify the key challenges in implementing moderation in decentralized systems.
     
    \item \emph{Moderation Mechanisms}. We propose two protocol-level moderation frameworks for Ethereum \IDMs: \BuildMod, where builders perform semantic checks during block construction, and \UserMod, where users proactively attach moderation proofs prior to transaction submission. These designs explore distinct trust and integration models for enforcing semantic validity on-chain. To the best of our knowledge, this is the first work to integrate content moderation design into Ethereum's transaction model.
    
    \item \emph{Evaluation}. We evaluate the \textit{per}-transaction and \textit{per}-block validation latency introduced by \BuildMod and \UserMod.  Our results show that while \BuildMod incurs significantly higher overhead due to in-block content classification, \UserMod remains lightweight and more practically viable under Ethereum’s tight block-building time constraints. We further propose and simulate a moderation fee mechanism design to incentivize builders to validate and include transactions with non-toxic content. Our simulation results show that, with appropriate parameter settings, \UserMod can achieve economic efficiency.
    
\end{itemize}

\section{Ethereum Fundamentals}
\label{sec: bg_ethereum}

\noindent\textit{Ethereum Transaction.} 
An Ethereum transaction is a cryptographically signed instruction that triggers a state change on the Ethereum blockchain~\cite{ivanov2023txt,buterin2013ethereum,wood2014ethereum}. It may involve the transfer of \ETH, the invocation of smart contract functions, or the deployment of new smart contracts~\cite{wood2014ethereum}.

\smallskip
\noindent\textit{Ethereum PBS.} 
\PBS~\cite{buterin2021pbs_fee_market} is a modular architecture that separates block proposing from building~\cite{heimbach2023ethereum}. Previously, validators handled both tasks, which concentrated \MEV opportunities~\cite{daian2020flash}. With \PBS, specialized builders construct blocks and submit bids; proposers simply choose the highest bid. This separation promotes more decentralized block production.

\smallskip
\noindent\textit{MEV-Boost.} 
\PBS is currently realized via \MEVBoost~\cite{flashbots-mev-boost}, an open-source middleware by Flashbots that enables an out-of-protocol auction for block construction. Builders construct and bid on blocks, relays serve as intermediaries that forward the most profitable bids, and proposers select among these bids to finalize block production (see Fig.~\ref{fig:mev_boost}). The builder retains \MEV profits and priority fees, while the proposer receives the payment specified in the winning bid.  As of today, around $90\%$ of Ethereum validators are utilizing \MEVBoost~\cite{mevboost_dashboard}.

\smallskip
\noindent\textit{\ePBS.} 
While \MEVBoost enables PBS via an off-chain market of builders and relays, it relies on trusted intermediaries to relay bids and deliver payloads. \ePBS, proposed in EIP-7732~\cite{eip7732}, moves key logic on-chain and eliminates the reliance by embedding bid submission and block building commitments directly into the consensus process.

\section{Input Data Messages (IDMs)}
Ethereum transaction includes an \inputdata field that is designed for encoding function calls to smart contracts. However, this field has been increasingly repurposed by users to embed natural language content in hexadecimal form~\cite{xiong2025talking}. These \IDMs are permanently stored on-chain and accessible to anyone as a form of decentralized communication.

\subsection{Use of IDMs for Toxic Messaging}
While many \IDMs are used for expressive purposes, Xiong~\etal~\cite{xiong2025talking} reveals that $4.3\%$ of the \IDMs (recorded between \StartTimeData and \EndTimeData) contain toxic content, including instances of profanity, threats, discriminatory language, sexually explicit content, and embedded links to hidden channels (see Tab.~\ref{tab:toxic_idm_data}). These toxic messages are directed at other users without any clear transactional context.

\begin{table}[htb]
\centering
\caption{\textbf{Breakdown of Toxic/Abusive IDMs} by subtopic, language, and frequency (data sourced from~\cite{xiong2025talking}).} 
\label{tab:toxic_idm_data}
\renewcommand\arraystretch{1.2}
\resizebox{\linewidth}{!}{
\begin{tabular}{c|c|cccc}
\toprule

\multicolumn{1}{c}{\multirow{2}{*}{\textbf{Subtopic}}} & \multicolumn{1}{c}{\multirow{2}{*}{\textbf{Example}}} & \multicolumn{2}{c}{\textbf{English}} & \multicolumn{2}{c}{\textbf{Chinese}} \\
\cmidrule(lr){3-4} \cmidrule(lr){5-6}
& & \textbf{\#Unique}  & \textbf{\#Total} & \textbf{\#Unique} & \textbf{\#Total} \\

\midrule

Verbal Abuse \& Profanity  & {0xd9...f83} & 1,961 & 2,582 & 97 & 103 \\
Threats, Harassment \& Psych & {0xa9...8e7}  & 215 & 268 & 116 & 116 \\
Discriminatory/Hate Speech & {0x6f...4b2} & 68 & 77 & 2 & 2 \\
Sexual/Pornographic Content & {0xbb...07c} & 36 & 50 & 32 & 32 \\
Hidden Channel Links & {0xb8...294} & 13 & 126 & 0 & 0 \\
\bottomrule
\end{tabular}
}
\begin{tablenotes}[flushleft]
      \footnotesize
      \item[] \quad \textit{IDM links have been omitted to avoid directing readers to harmful content.}
\end{tablenotes}
\end{table}

It is worth noting that a toxic rate of $4.3\%$ in Ethereum \IDMs is nontrivial, especially when contextualized against Web2 social media platforms. For example, Wang~\emph{et al.}~\cite{wang2014cursing} found that profanity occurred in only $1.15\%$ of tweets, while Dawkins~\emph{et al.}~\cite{dawkins2025detection} reported a slightly higher rate of $3.5\%$ in their dataset of human-written tweets. Regarding hate speech, Gao~\emph{et al.}~\cite{gao2017recognizing} observed that only $0.6\%$ of a randomly sampled set of $5{,}000$ tweets qualified as hateful content. These comparisons indicate that although decentralized communication through \IDMs is still in its early stages, the $4.3\%$ toxic rate is already substantial when measured against Web2 benchmarks.

\subsection{Impact of On-chain Toxic IDMs}

The persistent presence of toxic \IDMs on Ethereum calls for a closer look at their potential impact.

First, unlike Web2 messages that can be flagged, deleted, or downranked~\cite{singhal2023sok}, the on-chain nature of \IDMs means that once recorded, toxic messages are permanently stored and cannot be altered or removed. This allows harmful content to persist indefinitely and remain publicly accessible to all users.

Second, users have no means to block abusive senders or report harmful content, and there is no effective mechanism for moderation in decentralized systems. This lack of mitigation tools exacerbates the effects of targeted abuse and harassment. 

Third, the presence of toxic \IDMs can foster hostile environments that undermine trust and discourage participation in decentralized communities. For instance, 0x5b14...b26 and 0x4aC8...6c6\footnote{We removed address links to avoid directing readers to harmful content.} exchanged multiple insulting \IDMs back and forth. Such interactions show how unmoderated environments can escalate into sustained hostility between users.

\subsection{Challenges for On-chain Moderation of Toxic IDMs}

\subsubsection{\underline{Content Moderation in Web2}}

Content moderation on Web2 platforms aims to ensure safe, legal, and trustworthy environments. It typically involves three main components~\cite{singhal2023sok}.

\paragraph{Rules} The first step is rule definition, typically stated in a platform’s community guidelines or terms of service. Commonly restricted content includes copyright violations, hate speech, and misinformation~\cite{schaffner2024community}. Major platforms like Facebook, YouTube, and Twitter publicly list such rules~\cite{schaffner2024community}.

\paragraph{Detection} 
Content violations are identified through user reports, human review, or automated systems~\cite{gorwa2020algorithmic, singhal2023sok}. Human-in-the-loop methods, which integrate automation with human judgment, are also increasingly adopted~\cite{lykouris2024learning}.

\paragraph{Enforcement}
Content-targeted enforcement removes or demotes posts, while user-targeted enforcement includes suspensions or bans~\cite{singhal2023sok}. Platforms also apply softer measures like warnings or quarantines, especially for misinformation~\cite{singhal2023sok}. Enforcement practices, however, remain inconsistent and often lack transparency, particularly with automation~\cite{gorwa2020algorithmic, schaffner2024community}.

\smallskip
\subsubsection{\underline{Content Moderation in Web3}}
Advances in blockchain have enabled Web3 social media platforms that decentralize content ownership and governance.
One such platform is \textsf{memo.cash}, a microblogging service where all user interactions are stored immutably on the BCH blockchain~\cite{zuo2024understanding}.

Moderation in Web3 social media is fundamentally different from Web2. Without centralized platform authorities, moderation is often user-driven, relying on mechanisms such as personal block lists (e.g., user-level ``mute'' on \textsf{memo.cash})~\cite{zuo2023first,zuo2024understanding}. However, such a mechanism merely filters content at the individual level without affecting its availability or visibility to others. The absence of moderation mechanisms for Web3 social media can lead to persistent exposure to harmful or illegal material and limited collective accountability \cite{bovenzi2024content}.

\smallskip
\subsubsection{\underline{Challenges for On-chain Moderation}}
Immutability, decentralization, and pseudonymity in blockchain systems pose structural challenges to effective content moderation.
\begin{itemize}[leftmargin=*, topsep=2pt, itemsep=0pt]
    \item \emph{Immutability of content}: Once content is written to the blockchain, it cannot be altered or removed. This makes it impossible to remove harmful content after publication.
    \item \emph{Lack of centralized enforcement authority}: Web3 systems intentionally avoid central operators. As a result, there is no trusted party that can set policies and ensure enforcement. 
    \item \emph{Identity and accountability gaps}: Web3 emphasizes pseudonymity, which makes it difficult to assign responsibility or impose consequences for abusive behavior. Repeat offenders can easily re-enter the system under new identities.
\end{itemize}

\subsection{Legal Basis}
In Web2, content moderation is often a legal requirement. The EU’s Digital Services Act (DSA)~\cite{dsa2022}, for instance, mandates all platforms serving EU users to remove harmful content, provide user reporting channels, and publish transparency reports. Very large platforms face additional obligations, such as systemic risk assessments.
The UK’s Online Safety Act (OSA)~\cite{osa2023} imposes similar duties. It requires platforms to assess the risk of illegal or harmful content and take steps to reduce it. Moderation is a legal duty that applies to platforms that make user-generated content visible and accessible.

Decentralized communication platforms are not yet subject to formal regulation. However, as they are increasingly used for public messaging, concerns about oversight are growing. These platforms start to resemble traditional services that host user-generated content, but lack equivalent governance structures. As a result, regulators may begin to extend existing frameworks to include decentralized systems~\cite{friedl2024decentralised}. Anticipating this shift, Web3 platforms may need to adopt moderation mechanisms to manage legal risk and public accountability.

\section{System and Threat Model}

\subsection{System Model}
\label{sec: system_model}

\subsubsection{\underline{Ethereum Transaction}}
This work focuses on Ethereum transactions, which contain the following information~\cite{eth-tx}:

\begin{figure*}[tbh]
\centering
\includegraphics[width=0.99\linewidth]{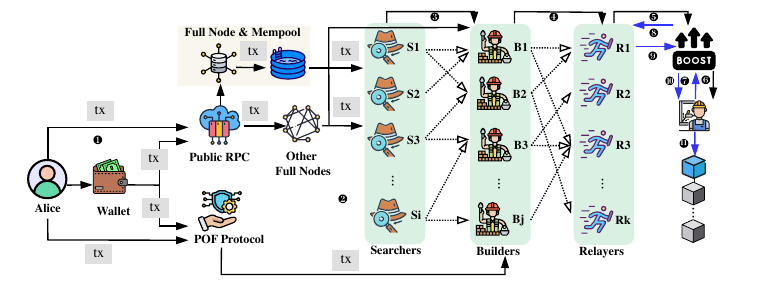}
    \caption{Overview of Alice's Ethereum transaction lifecycle under the \MEVBoost architecture.
    }
    \label{fig:mev_boost}
\end{figure*}

\begin{itemize}[leftmargin=*, topsep=2pt, itemsep=1pt]
    \item \textit{from}: The sender’s address that issues the transaction.
    \item \textit{to}: The recipient address, which is either an \EOA or a smart contract. Transactions to contracts trigger execution; those to \EOAs transfer value.
    \item \textit{signature}: A proof generated using the sender’s private key to authenticate the sender and ensure transaction integrity.
    \item \textit{nonce}: A counter associated with the sender’s account, indicating the number of \txs the sender has already issued.
    \item \textit{value}: The amount of \ETH to be sent to the recipient.
    \item \textit{gasLimit}: The upper bound on transaction gas consumption.
    \item \textit{maxPriorityFeePerGas}: The incentive paid directly to block producers to prioritize transaction validation and inclusion.
    \item \textit{maxFeePerGas}:  The maximum fee cap per gas unit.
    \item \textit{input data}: An optional byte array used to pass parameters or payloads, particularly when invoking smart contracts. In the context of decentralized communication, this field is often repurposed by senders to embed human-readable messages.
\end{itemize}

\smallskip
\subsubsection{\underline{Ethereum Transaction Lifecycle}}

An Ethereum transaction lifecycle typically involves the following processes: transaction generation, broadcasting, validation, inclusion, and finality~\cite{wang2023blockchain,pacheco2023my}. Fig.~\ref{fig:mev_boost} illustrates the transaction lifecycle under the current Ethereum \PBS implementation framework (i.e., \MEVBoost). The following steps describe this process in detail:
\protect\circnum{1} Alice initiates \tx either directly or via a wallet provider. \tx is routed either to a public RPC or to a \POF protocol. \protect\circnum{2} In the former case, \tx enters the mempool of the connected full node and is propagated to peers via gossip. Searchers monitor the public mempool for \MEV opportunities and construct bundles. Builders also monitor the mempool and may include such transactions directly when constructing blocks, either alongside or independently of bundled submissions. In the latter case, the \POF protocol sends \tx directly to participating builders.  \protect\circnum{3} Searchers submit bundles to builders for inclusion. Note that each searcher can submit the same bundle to multiple builders. \protect\circnum{4} Builders construct execution payloads and submit bids to relays. Note that each builder can submit the same bid to multiple relays. \protect\circnum{5} Each relay selects its highest-value bid and forwards the associated block header to \MEVBoost. \protect\circnum{6} \MEVBoost compares the received bids and delivers the highest one to the proposer. \protect\circnum{7} The proposer signs the selected block header and requests the full execution payload from \MEVBoost. \protect\circnum{8} \MEVBoost forwards the request to the originating relay. \protect\circnum{9} The relayer returns the full execution payload to \MEVBoost. \protect\circnum{10} \MEVBoost relays the payload to the proposer, who verifies its consistency with the signed header. \protect\circnum{11} Upon successful validation, the proposer assembles and broadcasts the full block to the network. Alice’s \tx is included on-chain if it is included in the execution payload of the winning bid.

\subsection{Threat Model}

We consider an adversary $\mathcal{A}$ capable of crafting arbitrary Ethereum transactions. Specifically, the adversary can inject semantically harmful content into the \inputdata field, thereby creating a toxic \IDM. Let a \tx be represented as:
{
\setlength{\abovedisplayskip}{6pt}
\setlength{\belowdisplayskip}{6pt}
\begin{equation}
\mathsf{tx} = (\mathsf{txFields}_{\text{meta}}, \mathsf{input})
\end{equation}
}

where $\mathsf{input} \in \mathbb{B}^*$ denotes an arbitrary-length byte string, with $\mathbb{B}^* = \{0,1\}^*$, i.e., sequences of arbitrary bytes. These strings may encode UTF-8 characters but are not restricted to valid UTF-8 encodings.
$\mathsf{txfields}_{\text{meta}}$ denotes all fields unrelated to semantic content, such as $\mathsf{to}$, $\mathsf{value}$, $\mathsf{gasLimit}$, etc.

\vspace{1mm}
The adversary then samples a toxic message $m_{\text{tox}} \in \mathcal{M}_{\text{tox}}$, where $\mathcal{M}_{\text{tox}}$ represents the space of semantically harmful messages (e.g., hate speech, threats, or illicit links). This message is encoded using the following public function:
{
\setlength{\abovedisplayskip}{6pt}
\setlength{\belowdisplayskip}{6pt}
\begin{equation}
\mathsf{input}_{\text{tox}} = \mathsf{encode}_{\mathsf{utf8}}(m_{\text{tox}})
\end{equation}
}

The adversary then constructs a \tx with the toxic payload:
{
\setlength{\abovedisplayskip}{6pt}
\setlength{\belowdisplayskip}{6pt}
\begin{equation}
\mathsf{tx}_{\text{tox}} = (\mathsf{txFields}_{\text{meta}}, \mathsf{input}_{\text{tox}})
\end{equation}
}

$\mathsf{tx}_{\text{tox}}$ is submitted to the network (e.g., via public RPC):
{
\setlength{\abovedisplayskip}{6pt}
\setlength{\belowdisplayskip}{6pt}
\begin{equation}
\mathcal{A} \rightarrow \textsf{eth\_sendRawTransaction}(\mathsf{tx}_{\text{tox}})
\end{equation}
}

Since Ethereum performs no semantic validation of input data, the transaction is valid and, once confirmed, will be included in a block $\mathsf{B}_i$ that is part of the canonical blockchain state $\mathcal{C}$, defined as a sequence of blocks:
{
\setlength{\abovedisplayskip}{6pt}
\setlength{\belowdisplayskip}{6pt}
\begin{equation}
\mathcal{C} = (\mathsf{B}_0, \mathsf{B}_2, \ldots, \mathsf{B}_n), \quad \mathsf{tx}_{\text{tox}} \in \mathsf{B}_i \text{ for some } i \in [0,n]
\end{equation}
}

\subsection{System Design Goals}\label{sec: goals}

\begin{itemize} [leftmargin=*, topsep=2pt, itemsep=1pt]
    \item \emph{Content Governance}: 
    Ethereum validates transactions syntactically and at the state level but lacks mechanisms for enforcing semantic properties, such as detecting harmful content in payloads. Our framework aims to introduce modular, programmable governance over transaction semantics.
    
    \item \emph{Security Preservation}: 
    Moderation must not compromise the blockchain’s core security guarantees, including safety, liveness, and censorship resistance for uninvolved users. It should avoid introducing vulnerabilities such as costly verification, proof manipulation, or selective denial of service.
    
    \item \emph{Economic Efficiency}: 
    Moderation should maintain the incentive compatibility and efficiency of Ethereum’s fee market. It must not disrupt the equilibrium of fee incentives for efficient block construction and MEV-aware inclusion.
\end{itemize}

\section{Design Framework}

This section presents two design frameworks for enabling content moderation on the Ethereum blockchain. We begin with \BuildMod, a builder-side moderation model inspired by Web2-style platform enforcement, where transaction semantics are analyzed during block construction. We then introduce \UserMod, a user-initiated scheme in which moderation occurs before transaction generation, and proof of moderation is embedded into the transaction. Each design reflects a different trust model and deployment tradeoff.

\subsection{\BuildMod: Builder-enforced inline moderation}
\subsubsection{\underline{Design Motivation}}
In Web2 platforms, moderation is typically executed by the platform itself, with minimal user involvement. This centralized enforcement model naturally prompts the question: In Ethereum’s decentralized communication environment, could moderation be similarly delegated to an infrastructural role? Within the \MEVBoost framework (see Fig.~\ref{fig:mev_boost}), builders are uniquely positioned for this task. Unlike proposers and relays, builders observe full transaction content and control inclusion at block construction time. 

\begin{figure}[tbh]
\centering
\includegraphics[width=0.99\linewidth]{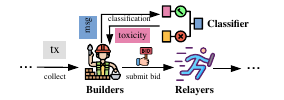}
    \caption{Overview of the \BuildMod framework.}
    \label{fig:builderMod}
\end{figure}

We therefore propose \BuildMod (see Fig.~\ref{fig:builderMod}), a moderation framework in which builders analyze the semantic content of transactions at block construction time. Transactions containing toxic content can be excluded, while valid transactions proceed through the standard execution pipeline. 

\subsubsection{\underline{System Components}}

\BuildMod introduces moderation at the builder level with the following entities:

\begin{itemize}[leftmargin=*, topsep=2pt, itemsep=1pt]
    \item \emph{User}: The transaction sender, who can compose a transaction that includes arbitrary \IDM in the \inputdata field.
    
    \item \textit{Builder}. Builders inspect transactions during block construction time. If the input contains semantic content,  builders apply moderation logic before deciding on inclusion.
    
    \item \textit{Classifier}. A moderation component accessible by the builder. It can be a local lightweight classifier embedded in the builder client, or a remote moderation service (e.g., via API) queried per transaction. Both variants will return binary results (e.g., \textit{toxic} or \textit{non-toxic}) via function $\mathsf{clf}(\cdot)$.
\end{itemize}

\vspace{1mm}
\subsubsection{\underline{Moderation Workflow}}

In \BuildMod, moderation is performed by builders during block construction (before step \circnum{4} in Fig.~\ref{fig:mev_boost}).  Let $\mathsf{tx}$ be a candidate transaction with message $\mathsf{input_{\text{msg}}}$ embedded in the \inputdata field. The builder first attempts to extract the message from the raw input:
{
\setlength{\abovedisplayskip}{6pt}
\setlength{\belowdisplayskip}{6pt}
\begin{equation}
\mathsf{msg} = \mathsf{decode}_{\mathsf{utf8}}(\mathsf{input_{\text{msg}}})
\label{eq:decode}
\end{equation}
}

If $\mathsf{msg}$ contains meaningful UTF-8 characters, i.e., $\mathsf{msg} \neq \bot$, the moderation procedure proceeds.  Here, $\bot$ denotes failure to decode valid semantic content.  Then the builder initiates a moderation check via a classifier function $\mathsf{clf}(\cdot)$ before including the transaction in the execution payload:
{
\setlength{\abovedisplayskip}{6pt}
\setlength{\belowdisplayskip}{6pt}
\begin{equation}
\mathsf{clf}(\mathsf{msg}) \in \{\text{toxic}, \text{non-toxic}\}
\label{eq:classify}
\end{equation}
}

Here we consider two types of moderation strategies:

\paragraph{Local Moderation} The builder runs a lightweight classifier locally within its block-building logic. This may range from simple keyword blacklisting (e.g., regex-based matching against profanity or hate terms), to natural language processing (NLP) models such as transformer-based binary classifiers optimized for latency. These lightweight methods offer fast inference, making them suitable for real-time block assembly. However, they tend to suffer from low recall and weak contextual understanding, and may be easily evaded by adversaries through obfuscation or phrasing.

\vspace{1mm}
\paragraph{External Classification} Alternatively, the builder can delegate moderation to an external classifier, such as a GPT-based large language model (LLM) hosted via API. The input data is submitted to this external service for toxicity classification. This approach benefits from higher semantic precision, broader coverage, and adaptive moderation logic. However, in high-load settings, reliance on external services may become infeasible due to rate limits. It may incur substantial latency that is often incompatible with the strict time budget builders face when competing in real-time \PBS environments. 

In both strategies, transactions without semantic content are processed through the standard execution pipeline without moderation. For transactions containing \IDMs, those identified as \textit{non-toxic} are accepted and considered for inclusion in the block, while those flagged as \textit{toxic} are excluded.

\vspace{1.5mm}
\subsubsection{\underline{Moderation Fee}}

To compensate builders for the cost of semantic moderation, we introduce a protocol-level fee, denoted \textsf{modExecFee}, that is paid by the sender of a transaction containing natural-language input data (i.e., \IDMs). 

\paragraph{Local Moderation} In this case, the classification cost is bounded and approximately constant per transaction. The fee is therefore modeled as a fixed constant:
\begin{equation}
\mathsf{modExecFee}_{\text{local}} = f_0 + f_1\cdot |\mathsf{input}_{\text{msg}}|
\end{equation}
where $f_0$ is a fixed base cost, and $f_1$ is a per-byte cost rate. 

\vspace{1mm}
\paragraph{External Classification} In this case, the builder outsources moderation to an external service that typically charges per-token pricing. We model the fee as:
\begin{equation}\label{eq: buildermod-fee}
\mathsf{modExecFee}_{\text{external}} = c \cdot \mathsf{numTokens}(\mathsf{input}_{\text{msg}})
\end{equation}
where $c$ is the per-token cost rate, and $\mathsf{numTokens}(\cdot)$ is the number of tokens under a tokenizer (e.g., \textsf{tiktoken}).

\vspace{1mm}
\paragraph{Fee Sufficiency Check} To ensure the moderation cost is sufficiently covered, each builder maintains a local policy to evaluate whether the fee provided by the user is sufficient for moderation execution. Transactions that fail to provide sufficient fees may be excluded during block construction.

\vspace{1mm}
\subsection{\UserMod: User-initiated pre-transactional moderation}

As discussed earlier, while \BuildMod aligns with Web2 moderation logic, it introduces practical limitations. Builders operate under strict time budgets to construct and submit competitive payloads. Executing per-transaction content classification, particularly when involving external services, can substantially increase latency and reduce responsiveness.  Therefore, we introduce an alternative design, \UserMod, which shifts moderation responsibility to the transaction sender.

\smallskip
\subsubsection{\underline{Design Motivation}}
The \inputdata field can be misused to enable users to embed immutable and publicly visible messages that may contain potentially harmful content.
In this context, the user is the only party with full knowledge of the semantic content of a transaction before submission. Moreover, the user should be responsible for the content they introduce into the system. This makes the user the most suitable actor to perform content checking. Pre-transactional moderation, carried out before the transaction is broadcast, enables users to assess the risk associated with the content. 

\begin{figure}[tbh]
\centering
\includegraphics[width=0.99\linewidth]{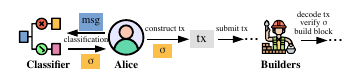}
    \caption{Overview of the \UserMod framework.}
    \label{fig:userMod}
\end{figure}

Therefore, we design the \UserMod framework (see Fig.~\ref{fig:userMod}) to formalize this user-initiated process. It requires users to prove that their transaction content is non-toxic, with the certificate from a pre-defined off-chain classifier. By shifting responsibility to the transaction sender, \UserMod aligns with the principle of end-user accountability without introducing computational overhead for other blockchain participants.

\smallskip
\subsubsection{\underline{System Components}}

\UserMod establishes a user-facing moderation framework that integrates semantic constraints into the Ethereum transaction lifecycle. The system is composed of five core components:

\begin{itemize}[leftmargin=*, topsep=2pt, itemsep=1pt]
    \item \emph{User}: The transaction sender, who can compose a transaction that includes arbitrary \IDM in the \inputdata field.
    
    \item \emph{Classifier}: An external moderation service (e.g., API-hosted LLMs) that evaluates whether a given \IDM contains toxic content and attests compliance with a digital signature.
    
    \item \emph{Moderation Proof}: A cryptographic certificate attesting that the associated \inputdata has passed moderation.
    
    \item \emph{Transaction with Embedded Proof}: A transaction where the user appends the moderation proof to the end of the \inputdata field before submission.
    
    \item \emph{Builder}: A block builder that inspects transactions and verifies moderation proofs before inclusion.
\end{itemize}

\smallskip
\subsubsection{\underline{Moderation Workflow}} 
\UserMod enforces semantic moderation at the user side before a transaction submission (before step \circnum{1} in Fig.~\ref{fig:mev_boost}). It involves three stages (see Fig.~\ref{fig:userMod}):

\vspace{0.5mm}
\paragraph{User-side Moderation and Transaction Construction}
In the \UserMod model, if the user embeds human-readable messages within \inputdata, denoted $\mathsf{input}_{\text{msg}}$, this content must be semantically moderated prior to submission. The user sends $\mathsf{input}_{\text{msg}}$ to a trusted classifier. Upon successful moderation, the classifier returns a signature:
{
\setlength{\abovedisplayskip}{6pt}
\setlength{\belowdisplayskip}{6pt}
\begin{equation}
\sigma = \mathsf{sign}_{\mathsf{sk}_{\text{clf}}}(H(\mathsf{input}_{\text{msg}}))
\end{equation}
}

The final input to be embedded in the \tx is constructed as:
{
\setlength{\abovedisplayskip}{6pt}
\setlength{\belowdisplayskip}{6pt}
\begin{equation}
\mathsf{input}_{\text{final}} = \mathsf{input}_{\text{msg}} \,\|\, \sigma
\end{equation}
}

This new input replaces the original one in the transaction. The moderated transaction $\mathsf{tx}_{\text{mod}}$ is thus defined as:
{
\setlength{\abovedisplayskip}{6pt}
\setlength{\belowdisplayskip}{6pt}
\begin{equation}
\mathsf{tx}_{\text{mod}} = (\mathsf{txFields}_{\text{meta}},\; \mathsf{input}_{\text{final}})
\end{equation}
}

The user then submits $\mathsf{tx}_{\text{mod}}$ through the network.

\vspace{0.5mm}
\paragraph{Transaction Submission}

$\mathsf{tx}_{\text{mod}}$ is submitted either via a public RPC endpoint into the mempool or through a \POF channel directly to connected builders (see Fig.~\ref{fig:mev_boost}). Both methods are compatible with \UserMod and preserve the ability for builders to verify the moderation proof.

\vspace{0.5mm}
\paragraph{{Builder-side Verification and Inclusion}}
Upon receiving a transaction, the builder decodes $\mathsf{input}_{\text{tx}}$ to determine whether it contains semantic content. It then applies a filtering policy based on the presence and validity of the attached moderation proof $\sigma$. Importantly, builders construct execution payloads incrementally by executing transactions sequentially and tracking state transitions. Thus, each included $\mathsf{tx}$ must be \emph{executable-valid} in terms of gas, nonce, and signature, and must also satisfy the \emph{semantic validity} defined by our moderation model. Formally, we define \emph{semantic validity} as:
\begin{equation}
{\small
\mathsf{semValid}(\mathsf{tx}) =
\begin{cases}
\text{true,} \quad \text{if } \mathsf{decode}_{\mathsf{utf8}}(\mathsf{input}_{\text{msg}}) = \bot \\
\mathsf{verify}_{\mathsf{pk}_{\text{clf}}}(\sigma, H(\mathsf{input}_{\text{msg}})) = \text{true,}\quad \text{otherwise}
\end{cases}
}
\end{equation}
This means that:
\begin{itemize}[leftmargin=*, itemsep=2pt]
    \item If no semantic content is detected, $\mathsf{semValid}$ is true.
    \item If semantic content is present and a valid proof $\sigma$ is attached (i.e., $\mathsf{verify}_{\mathsf{pk}_{\text{clf}}}(\sigma, H(\mathsf{input}_{\text{msg}})) = \text{true}$), $\mathsf{semValid}$ is true.
    \item If semantic content is present but $\sigma$ is missing or invalid (i.e., $\mathsf{verify}_{\mathsf{pk}_{\text{clf}}}(\sigma, H(\mathsf{input}_{\text{msg}})) = \text{false}$), $\mathsf{semValid}$ is false.
\end{itemize}

\smallskip
A transaction is included in the payload if and only if it is executable-valid and semantic-valid:
{
\setlength{\abovedisplayskip}{6pt}
\setlength{\belowdisplayskip}{6pt}
\begin{equation}
\mathsf{isValid}(\mathsf{tx}) = \mathsf{execValid}(\mathsf{tx}) \land \mathsf{semValid}(\mathsf{tx})
\end{equation}
}

Transactions failing either condition are excluded prior to payload submission. This workflow establishes a moderation-aware filtering layer while preserving execution semantics.

\smallskip
\subsubsection{\underline{Builder Incentive Fee}}

In \UserMod, each transaction carrying semantic content must embed a moderation proof, a digital signature generated by an external classifier certifying the non-toxicity of the message. Upon inclusion, the builder is responsible for verifying this signature to ensure semantic validity.
This verification imposes additional computational overhead compared to standard Ethereum transaction processing. Specifically, while a standard ETH transfer requires a fixed gas usage of $21{,}000$, moderation-aware transactions incur extra cost due to two additional operations: \textit{(1)} hashing the semantic message, and \textit{(2)} verifying the classifier's signature. Since the cost of hashing scales approximately linearly with the input size, we model the total gas usage as:
\begin{equation}\label{eq: usermod-gas}
\mathsf{gasUsed}_{\text{mod}} = \mathsf{baseGas} + \alpha + \beta \cdot |\mathsf{input}_{\text{msg}}|
\end{equation}

where $\mathsf{baseGas} = 21{,}000$ is the baseline for a simple ETH transfer; $\alpha$ is a constant capturing the fixed cost of signature verification (e.g., one \textsf{ecrecover}); $\beta$ represents the per-byte cost of computing the hash over the message; $|\mathsf{input}_{\text{msg}}|$ is the byte-length of the embedded semantic message. As a result, the total fee paid by the user becomes:
\begin{equation}
\mathsf{fee} = \mathsf{gasUsed}_{\text{mod}} \cdot (\mathsf{baseFee} + \mathsf{priorityFee})
\end{equation}

The builder receives the following protocol-defined reward:
\begin{equation}
\mathsf{builderReward} = \mathsf{gasUsed}_{\text{mod}} \cdot \mathsf{priorityFee}
\end{equation}

This formulation ensures that moderation becomes a gas-metered computation path and the builder is naturally incentivized to perform moderation-proof verification. 
{In practice, to realize this incentive mechanism, moderation-aware users must explicitly raise the \textsf{gasLimit} field of their transaction beyond the default baseline. For example, in a standard ETH transfer, wallets or clients typically set $\mathsf{gasLimit} = 21{,}000$. In our model, moderation-aware transactions must provision additional headroom to account for the verification cost:}
\begin{equation}\label{eq: user-gas-limit}
\mathsf{gasLimit} = \mathsf{baseGas} + \alpha + \beta \cdot |\mathsf{input}_{\text{msg}}| + \varepsilon
\end{equation}
where $\varepsilon$ provides a small buffer to avoid out-of-gas reverts. Wallets can automate this adjustment based on the message length and known verification complexity. Builders, in turn, will execute and include only those transactions that allocate sufficient gas and pass the moderation verification logic.

\section{Evaluation}

\subsection{Setup}

We modify the \href{https://github.com/ethereum/go-ethereum}{\textsf{go-ethereum}} client to enable input data moderation in a private Ethereum network. A moderation routine is added to the \textsf{txpool} validation logic, using LLMs from \href{https://openrouter.ai/}{\textsf{openrouter.ai}} to classify toxic \IDMs.

In the \BuildMod setup, the moderation function decodes input data from hexadecimal to UTF-8 and verifies whether the result is meaningful text. If valid, the decoded text is sent to the LLM classifier to determine if it contains toxic content.  In the \UserMod setup, the transaction sender queries the LLM to confirm that the input is non-toxic. The LLM’s cryptographic signature is then appended to the \inputdata, which is verified during the transaction validation process.

Experiments are conducted on an Apple MacBook Pro with an Apple M4 Max chip, featuring a 16-core CPU (12 performance cores and 4 efficiency cores), 128 GB of unified memory, and 4 TB of SSD storage, running on Sequoia v15.5.

\subsection{Validation Time Evaluation}
\subsubsection{\underline{Per-Transaction Validation Time}}

\begin{figure}[t]
\centering
\includegraphics[width=0.99\linewidth]{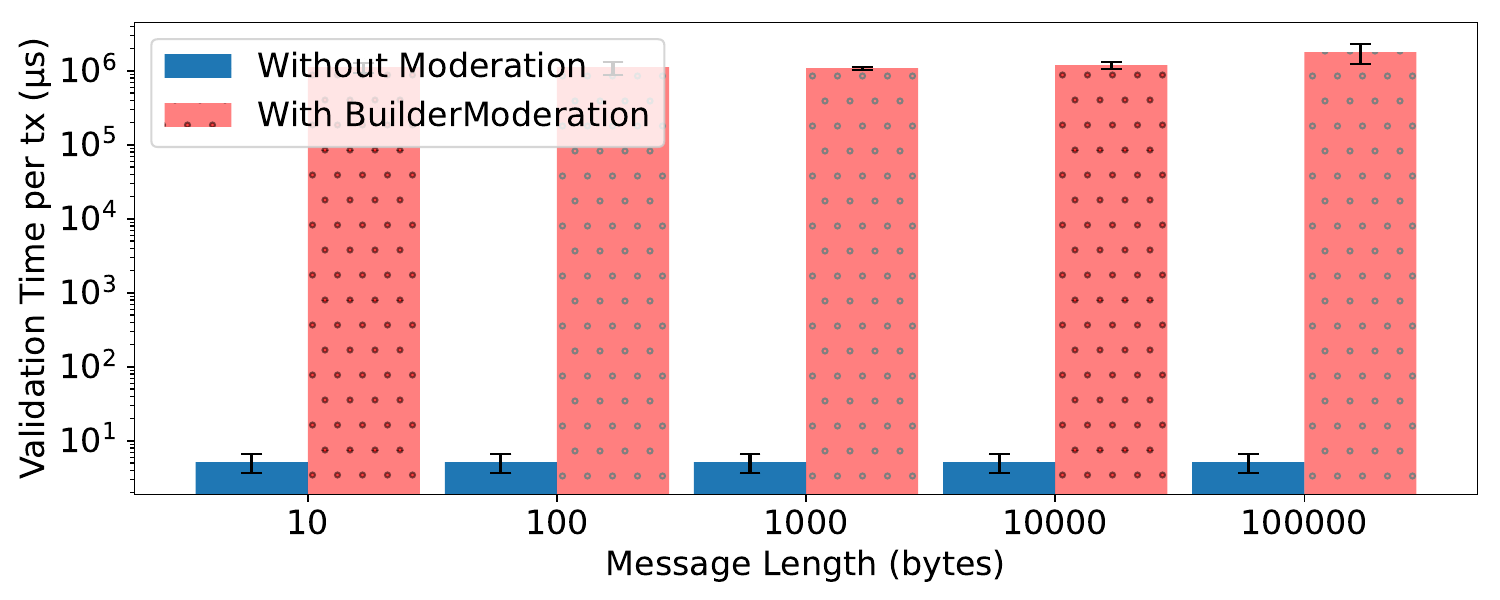}
    \caption{Comparison of \emph{per-transaction} validation time (\us) without and with moderation (\BuildMod). Results are based on outputs from the \textsf{openai/gpt-4.1} model.
    }
    \label{fig:buildmod_validation_time_comparison}
\end{figure}

\begin{figure}[t]
\centering
\includegraphics[width=0.99\linewidth]{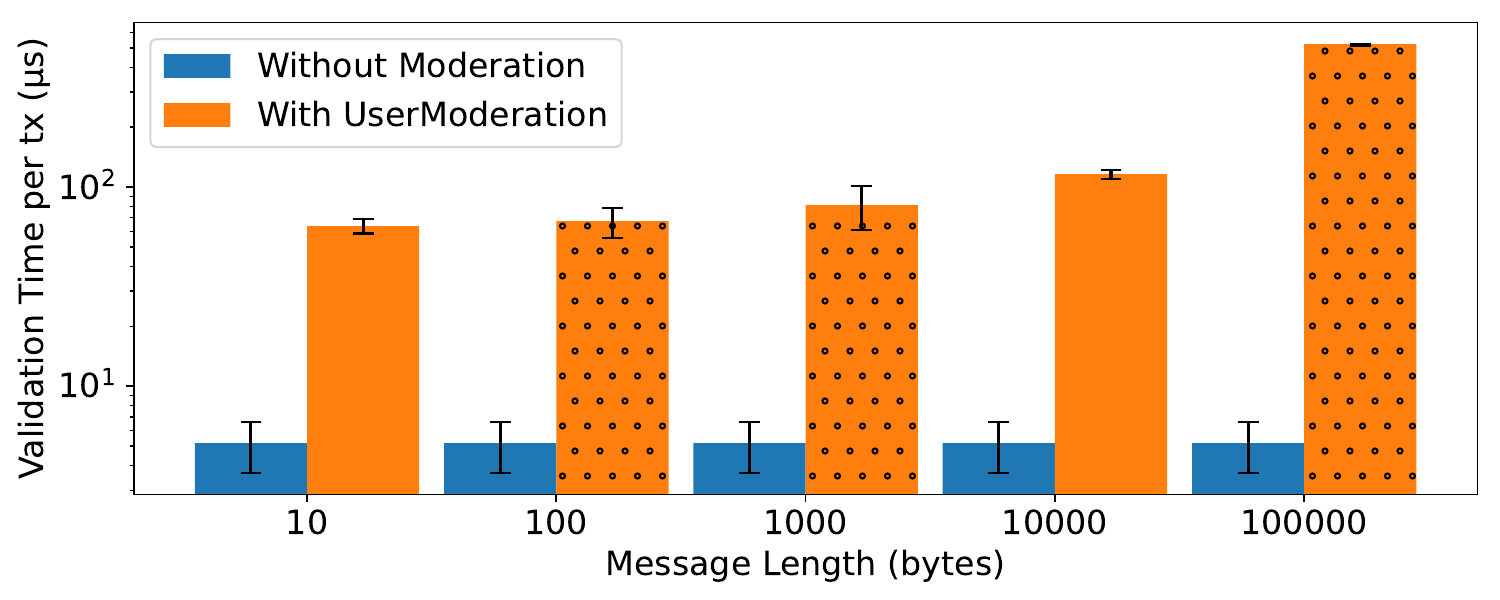}
    \caption{Comparison of \emph{per-transaction} validation time (\us) without and with moderation (\UserMod). Results are based on outputs from the \textsf{openai/gpt-4.1} model.
    }
    \label{fig:usermod_validation_time_comparison}
\end{figure}

We first evaluate the per-transaction validation time under different moderation regimes. As a baseline, we simulate the standard Ethereum transaction validation pipeline without moderation. This yields an average validation time of $5.12$~\us per transaction (shown as the blue bars in Fig.~\ref{fig:buildmod_validation_time_comparison} and Fig.~\ref{fig:usermod_validation_time_comparison})

We then measure the additional validation overhead introduced by content moderation. In the \BuildMod setting, builders are responsible for performing real-time toxicity classification either locally (e.g., lightweight models) or via external services (e.g., API-hosted LLMs). In this evaluation, we simulate an external moderation setup using the \textsf{openai/gpt-4.1} language model. As illustrated in Fig.~\ref{fig:buildmod_validation_time_comparison}, the moderation process introduces substantial latency. For instance, at a message length of $100$ bytes, a threshold below which $80\%$ of \IDMs fall as reported by~\cite{xiong2025talking}, the validation time increases from $5.12$~\us to $1{,}098{,}690.91$~\us. For extremely long messages ($10^5$ bytes), the latency reaches $1{,}772{,}946.65$~\us.

By contrast, in the \UserMod model, moderation is performed pre-submission by the user, and the builder only verifies a digital signature embedded in the transaction input. As shown in Fig.~\ref{fig:usermod_validation_time_comparison}, this drastically reduces validation time. At the same 100-byte threshold, \UserMod incurs only $66.94$~\us validation time on average, five orders of magnitude faster than that of \BuildMod. This demonstrates that \UserMod can offer lower latency and greater scalability in moderation-aware Ethereum environments, especially under high-throughput or time-sensitive block-building scenarios.

\smallskip
\subsubsection{\underline{Per-Block Validation Time}}

While per-transaction validation latency quantifies the overhead introduced by moderation at a fine granularity, its actual impact at the block level requires further investigation, as builders must validate hundreds of transactions under strict time constraints. We therefore evaluate the average per-block validation time. 

Following existing work~\cite{xiong2025talking}, we observe $5{,}238{,}336$ transactions with decodable UTF-8 content from Ethereum block \StartBlockData (\StartTimeData) to block \EndBlockData (\EndTimeData). During the same period, we observe a total number of $227{,}4470{,}454$ transactions from \href{https://etherscan.io/chart/tx}{Etherscan}. Therefore, for each block, the average number of transactions is $\frac{227{,}4470{,}454}{19{,}314{,}987-0} = 117.78 $ and the average number of transactions with \IDMs is $\frac{5{,}238{,}336}{19{,}314{,}987-0} = 0.27$.

\begin{figure}[t]
\centering
\includegraphics[width=0.99\linewidth]{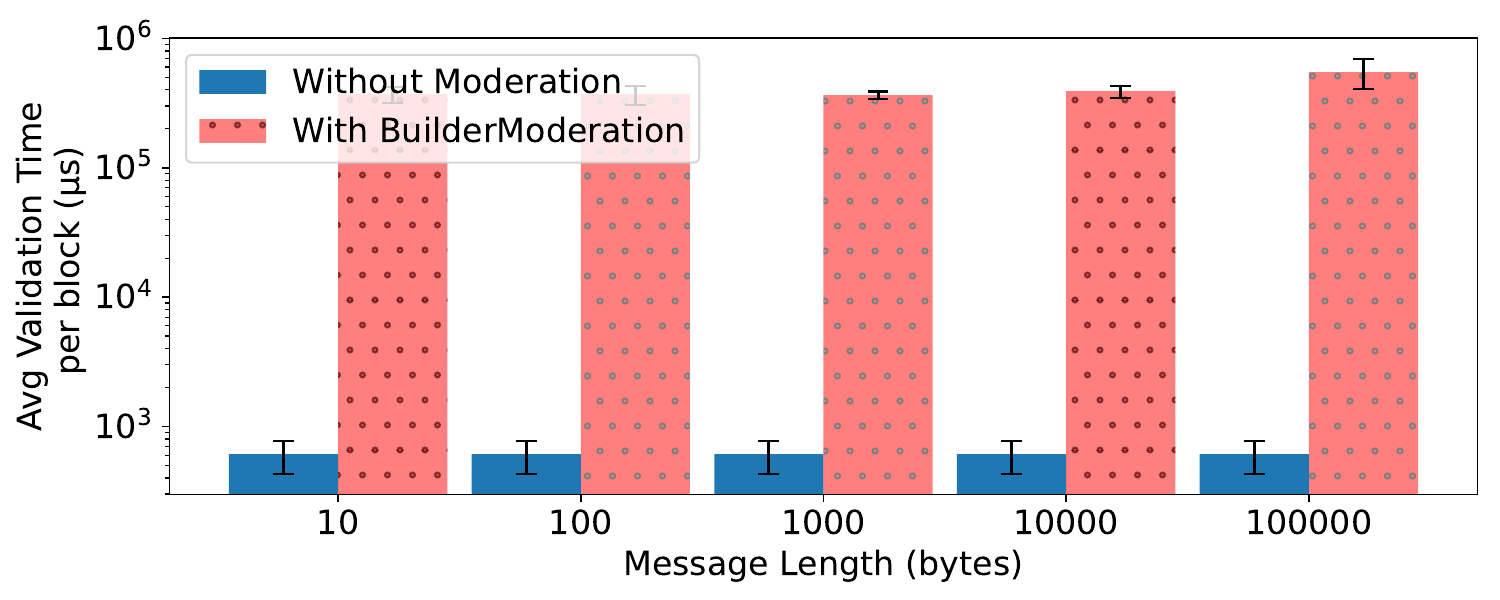}
    \caption{Comparison of average \emph{per-block} validation time (\us) without and with moderation (\BuildMod). Results are based on outputs from the \textsf{openai/gpt-4.1} model.
    }
    \label{fig:buildmod_validation_time_comparison_per_block}
\end{figure}

\begin{figure}[t]
\centering
\includegraphics[width=0.99\linewidth]{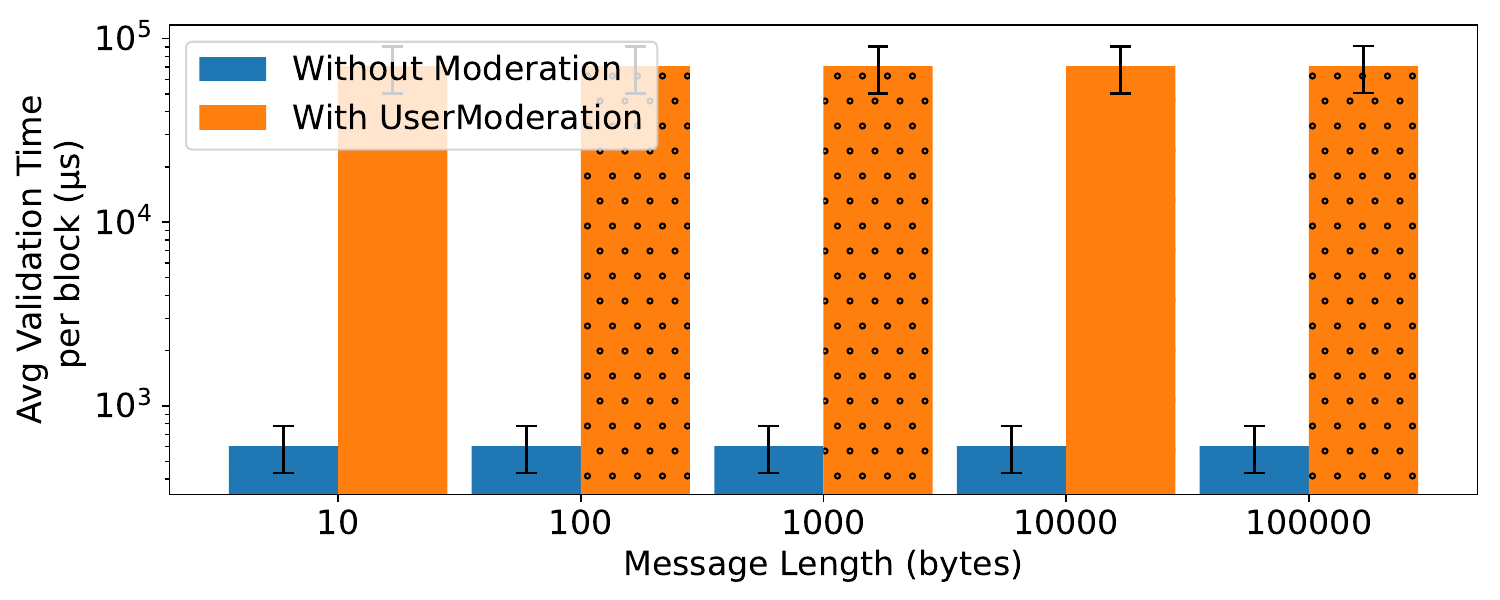}
    \caption{Comparison of average \emph{per-block} validation time (\us) without and with moderation (\UserMod). Results are based on outputs from the \textsf{openai/gpt-4.1} model.
    }
    \label{fig:usermod_validation_time_comparison_per_block}
\end{figure}

Using this \IDM density, we compute the expected validation time per block under both moderation models. The results are shown in Fig.~\ref{fig:buildmod_validation_time_comparison_per_block} and Fig.~\ref{fig:usermod_validation_time_comparison_per_block}. At the critical message length threshold of $100$ bytes, we observe that without moderation, both settings incur a baseline block-level validation time of $602.86$~\us. When moderation is enabled, \BuildMod incurs a substantially higher block validation cost of $367{,}048.33$~\us, due to the inference latency of external classifiers. In contrast, \UserMod only incurs $70{,}419.86$~\us validation time.

\vspace{0.1mm}
Overall, our evaluation results of validation time per-transaction and per-block both show the advantages of the \UserMod over \BuildMod, especially when processing transactions under stringent latency constraints.

\subsection{Cost Evaluation}

To assess the financial overhead of integrating third-party language models into the moderation pipeline, we evaluate the per-query cost of classifying a fixed-length \IDM using several publicly available API-hosted LLMs. Tab.~\ref{tab: llm_comparision} presents the moderation latency, tokenization results, and cost (in USD) incurred for processing an input \IDM of $1{,}000$ bytes.

\begin{table}[tbh]
\centering
\caption{\textbf{Moderation costs} comparison across LLMs for a fixed-length input (IDM length of $1{,}000$ Bytes).}
\label{tab: llm_comparision}
\renewcommand\arraystretch{1.1}
\resizebox{\columnwidth}{!}{
\begin{tabular}{l|lcl}
\toprule
\multicolumn{1}{c}{\textbf{Model}} & \multicolumn{1}{c}{\textbf{Time} (s)} & \textbf{Tokens} & \textbf{Costs} (USD) \\
\midrule
\textsf{openai/gpt-4.1} & 0.616 $\pm$ 0.117 &341
 &0.00069 \\
\textsf{openai/gpt-4o} & 0.921 $\pm$ 0.332 &341 &0.000863 \\
\textsf{qwen/qwen3-coder} & 1.214 $\pm$ 0.449 &354 &0.000715 \\
\textsf{google/gemini-2.5-pro} & 6.649 $\pm$ 0.717 &333 &0.00445 \\
\textsf{google/gemma-3-27b-it} & 0.901 $\pm$ 0.381 &338 &0.0000347 \\
\textsf{anthropic/claude-sonnet-4} & 13.319 $\pm$ 4.091 &420 &0.00132 \\
\textsf{deepseek/deepseek-r1-0528} & 7.463 $\pm$ 2.812 &347 &0.000207 \\
\bottomrule
\end{tabular}
}
\end{table}

Although the input content remains identical across all models, the number of resulting tokens varies due to differences in tokenizer implementations, which directly impacts pricing. For example, OpenAI's \textsf{gpt-4.1} tokenizes the message into $341$ tokens with a mean latency of $0.616$ seconds and a cost of $0.00069$, while Google's \textsf{gemma-3-27b-it} yields $338$ tokens at significantly lower cost ($0.0000347$). Notably, models such as \textsf{claude-sonnet-4} and \textsf{deepseek-r1-0528} exhibit substantially longer inference times, which may affect suitability for time-constrained builder operations.

\subsection{Transaction Fee Simulation}

To offset the computational costs caused by the moderation process, a transaction \IDM sender is required to pay additional fees to block validators (e.g., builders).

For \BuildMod, we estimate these fees via Eq.~\ref{eq: buildermod-fee}. As shown in Tab.~\ref{tab: llm_comparision}, the cost of querying LLMs is relatively low. For example, the most expensive model, Google’s \textsf{gemini-2.5-pro}, charges only $0.00445$ USD for processing a 1000-byte \IDM. However, the latency introduced by the \BuildMod computation can reduce builders’ ability to capture MEV opportunities and may lead to potential losses. Consequently, it is challenging to estimate the fees the message sender should pay sufficiently to compensate block builders.

\begin{figure}[tbh]
\centering
\includegraphics[width=0.95\linewidth]{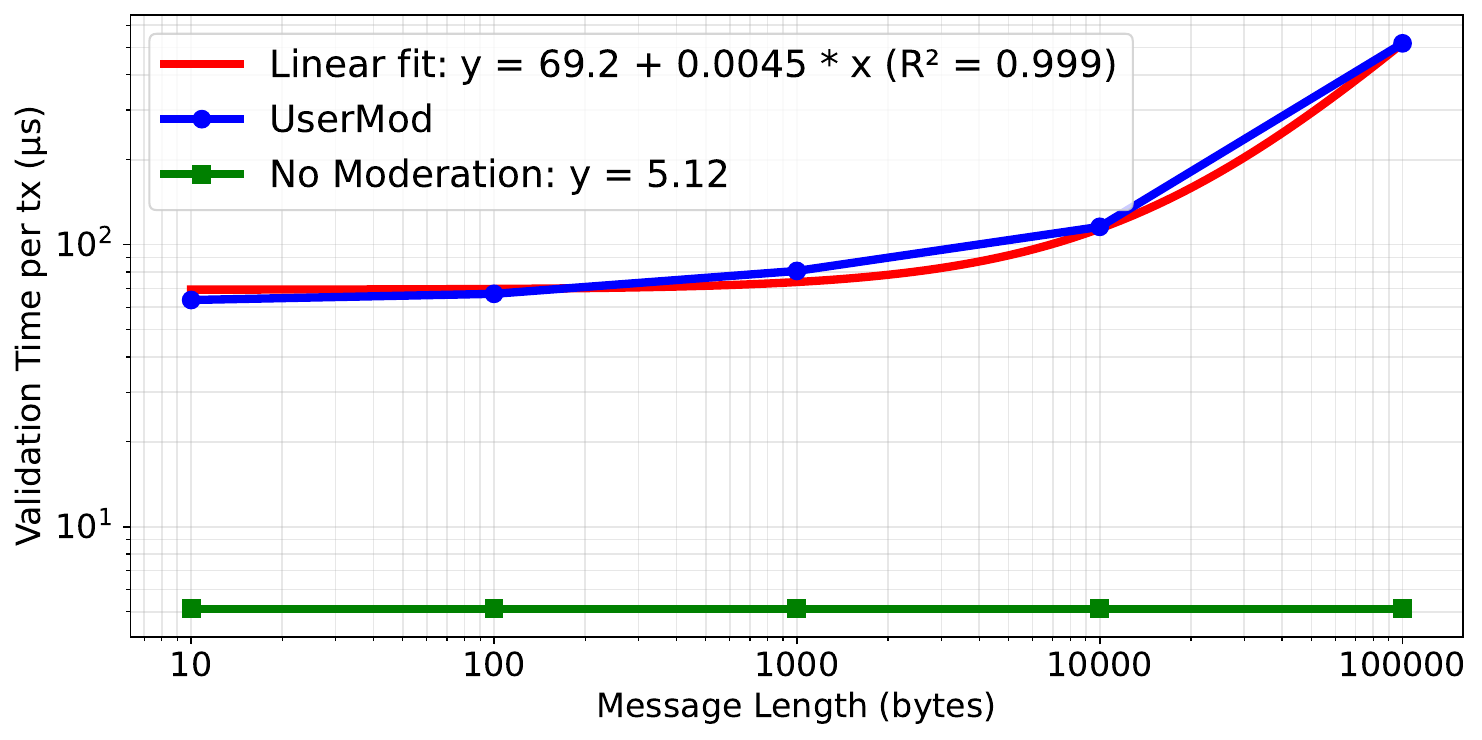}
    \caption{Time complexity of \UserMod validation (per \tx).
    }
    \label{fig:usermod_validation_time_single_tx}
\end{figure}

To simulate the moderation-related gas fees for \UserMod, we analyze the execution time with and without \UserMod to estimate the additional gas consumption introduced by moderation. As shown in Fig.~\ref{fig:usermod_validation_time_single_tx}, the execution time caused by \UserMod can be modeled by the linear function $y = 69.2 + 0.0045 \cdot x$, where $x = \mathsf{input}_{\text{msg}}$ denotes the length of the \IDMs. In contrast, the baseline validation time without moderation is $5.12$~$\mu$s, corresponding to the basic gas limit $ \mathsf{baseGas} = 21{,}000$. By incorporating Eq.~\ref{eq: usermod-gas}, we can estimate the gas usage for \UserMod as:
\begin{equation}\label{eq:usermod-gas-simulated}
{\small
\begin{aligned}
\mathsf{gasUsed}_{\text{mod}} 
&= 21000 + (69.2 - 5.12) \times \frac{21000}{5.12} \\
&\quad + 0.0045 \times \frac{21000}{5.12} \cdot |\mathsf{input}_{\text{msg}}|\\
& \approx 21000 + 262828 + 4102 \cdot |\mathsf{input}_{\text{msg}}|
\end{aligned}
}
\end{equation}

\section{Discussion}

This section discusses whether \BuildMod and \UserMod can achieve the three system design goals  (see \S\ref{sec: goals}).

In \BuildMod, builders need to query the LLM classifiers to determine if a transaction contains toxic content. Assuming the LLM classifiers are trusted, this design can fulfill the requirement of \emph{content governance}. And builders can select the transactions with sufficient fees to validate, mitigating DoS attacks. Therefore, \BuildMod can also guarantee \emph{security preservation}. However, it fails to achieve \emph{economic efficiency}, as the reliance on LLM queries introduces significant delays in block validation (see Fig.~\ref{fig:buildmod_validation_time_comparison} and Fig.~\ref{fig:buildmod_validation_time_comparison_per_block}), potentially resulting in missed MEV opportunities.

\UserMod can similarly ensure both \emph{content governance} and \emph{security preservation}, for the same reasons outlined in \BuildMod. Furthermore, in \UserMod, transaction senders are responsible for generating proofs of non-toxicity for \IDMs. As content verification and approval occur prior to transaction submission, this process imposes no additional LLM-caused delay on block builders. In addition, to incentivize builders to verify moderation proofs and include the transactions with \IDMs, senders must increase the gas limit (see Eq.~\ref{eq: user-gas-limit}). With proper parameter settings (see Fig.~\ref{fig:usermod_validation_time_single_tx} and Eq.~\ref{eq:usermod-gas-simulated}), \UserMod can also achieve \emph{economic efficiency}.

\section{Ethical Considerations}

\noindent\textit{Data Sensitivity}. 
This study analyzes toxic \IDMs. To reduce exposure risks, we omit full transaction links, abbreviate addresses, and redact sensitive text when necessary.

\vspace{0.05mm}
\noindent\textit{Research Intent}. 
This analysis is for research purposes only. Examining toxic content does not imply endorsement. The goal is to understand its implications for future moderation.

\vspace{0.05mm}
\noindent\textit{Data Collection}. 
Our study relies on publicly available data from the Ethereum blockchain. We made no attempt to deanonymize users or link addresses to real-world identities.

\vspace{0.05mm}
\noindent\textit{LLM Involvement}. In our moderation framework design, we use LLMs for toxicity classification.  While this enables scalable analysis, the model's classifications may reflect bias or produce errors. We interpret results with this limitation in mind and do not treat model outputs as definitive labels.

\section{Related Work}

\noindent\textit{Decentralized Social Media}. 
Driven by advances in blockchain technologies, Web3 has garnered increasing attention in recent years~\cite{wang2022exploring,yu2023leveraging}, fostering the emergence of decentralized social media platforms such as \textsf{memo.cash} and \textsf{Steemit}~\cite{zuo2023first,zuo2024understanding,zuo2023set,li2019incentivized}. Ethereum further supports embedding human-readable messages via the \inputdata field, known as \IDMs~\cite{xiong2025talking}. Beyond these, researchers have explored diverse desings for decentralized social systems~\cite{prodan2019articonf,prashanth2023socialchain}.

\vspace{1mm}
\noindent \textit{LLM-assisted Content Moderation}. As online communication grows in scale and complexity, traditional moderation via human review and user reports faces significant limitations~\cite{singhal2023sok}. Large Language Models (LLMs) have thus been explored as scalable and context-aware moderation tools to detect harmful content such as hate speech and offensive language~\cite{chiu2021detecting,franco2023analyzing,yang2023hare,huang2023chatgpt,oliveira2023good,kolla2024llm,kumar2024watch,zeng2024shieldgemma,li2024hot,yin2025bingoguard}.

\vspace{1mm}
\noindent \textit{Content Moderation in Decentralized Systems}.
Content moderation in decentralized systems faces inherent challenges. For instance, Hassan~\etal~\cite{hassan2021exploring} study moderation in the federated platform \textsf{Pleroma}, highlighting collateral damage from coarse-grained rejection. In blockchain-based systems, immutability further complicates moderation. Zuo~\etal~\cite{zuo2023first,zuo2024understanding,zuo2023set} examine \textsf{memo.cash} and its ``mute'' primitive, which hides rather than removes toxic content. Their analysis suggests muting is often ineffective due to high costs and its correlation with user activity rather than content toxicity.

Unlike \textsf{memo.cash}, Ethereum \IDMs are not part of a standardized social protocol, but serve as auxiliary metadata for informal communication. Xiong~\etal~\cite{xiong2025talking} highlight both the misuse of Ethereum \IDMs and the lack of content moderation frameworks. To address the gap, our work aims to initiate a design space for protocol-level content governance.

\section{Limitations and Future Work}

First, the effectiveness of moderation depends on the reliability of the classifier. While LLMs perform well in detecting toxicity, results may vary due to versioning, prompt sensitivity, or adversarial inputs. Future work should explore robust strategies such as confidence-aware checks or user appeals.

Second, our framework builds on the current \textsf{mev-boost} builder-relay model, but Ethereum is transitioning toward \ePBS (see \S\ref{sec: bg_ethereum}). This shift raises questions about integrating moderation models into protocol-native \PBS. Future work may adapt the framework to evolving \PBS designs.

Third, semantic moderation raises governance challenges due to varying and evolving legal definitions (e.g., of ``illegal content''). While our design supports flexible classifier-level policies, aligning them with jurisdictional standards remains unresolved. Future systems may require legally pluggable modules, auditable classification logs, and decentralized mechanisms for policy adaptation under legal guidance.

\section{Conclusion}

This paper presents a systematic exploration of protocol-level content moderation for Ethereum \IDMs. Motivated by the increasing misuse of \IDMs, we identify the semantic risks introduced by immutable and unmoderated communication. Given the lack of protocol-native mechanisms to moderate semantic misuse of \IDMs, we propose two frameworks: \BuildMod and \UserMod. Through evaluation and incentive modeling, we demonstrate that content moderation via \UserMod can be integrated into the transaction lifecycle. 

Looking ahead, a more robust and accountable on-chain communication environment will require advances in moderation classifiers, adaptation to execution-layer changes such as ePBS, and closer alignment with regulatory developments.  We hope this study motivates continued exploration into semantic governance as a foundational layer of decentralized systems.

\bibliographystyle{ieeetr}
\bibliography{references}

\end{document}